\documentclass[]{article}
\usepackage{graphicx}
\usepackage{wrapfig}
\usepackage{fullpage}

\long\def\comment#1{}
\newcommand{\commentout}[1]{}

\newcommand{\source}[1]{\textsf{#1}}

\newcommand{\secref}[1]{Section~\ref{#1}}
\newcommand{\figref}[1]{Figure~\ref{#1}}
\newcommand{\tabref}[1]{Table~\ref{#1}}

\newcommand{\denselist}{
      \setlength{\itemsep}{0pt}
      \setlength{\parsep}{1.5pt}
      \setlength{\topsep}{1.5pt}
      \setlength{\parskip}{2pt}
      \setlength{\partopsep}{0pt}
      \setlength{\labelwidth}{0em}
      \setlength{\labelsep}{0.5em} }

\newcommand{\bdesc}{\begin{description}\denselist}
\newcommand{\edesc}{\end{description}}

\begin{document}

\title{Social Browsing \& Information Filtering in Social Media}
\author{Kristina Lerman\\
University of Southern California \\
Information Sciences Institute\\
4676 Admiralty Way\\
Marina del Rey, California 90292, USA\\
lerman@isi.edu}

\maketitle

\begin{abstract}
Social networks are a prominent feature of many social media sites,
a new generation of Web sites that
allow users to create and share content.
Sites such as Digg, Flickr, and Del.icio.us allow users to designate others as
``friends'' or ``contacts'' and provide a single-click interface to track friends' activity.
How are these social networks
used? Unlike pure social networking sites (e.g., LinkedIn and Facebook), which allow
users to articulate their offline professional and personal
relationships, social media sites are not, for the most part, aimed at helping
users create or foster offline relationships.
Instead, we claim that social media users
create social networks to express their tastes and interests,
and use them to filter the vast stream of new
submissions to find interesting content.
Social networks, in fact, facilitate new ways of
interacting with information: what we call
\emph{social browsing}. Through an
extensive analysis of data from Digg and Flickr,
we show that social browsing is one of the primary usage modalities on these social media sites.
This finding has implications for how social media sites rate
and personalize content.

\end{abstract}


\section{Introduction}

The new 'social media' sites are changing the way Web users
interact with information. Unlike traditional Web sites,
Flickr, Del.icio.us, Digg, and YouTube, among others, enable
users to create and share information, knowledge, and opinions.
Social media sites share
the following elements: (1) users create or contribute
content, (2) users annotate content with descriptive keywords, or
tags, (3) users evaluate content, either passively or actively,
and (4) users create social networks by designating other users
with similar interests as \emph{friends} or \emph{contacts}.
Social media sites are to be distinguished from the explicit or pure social networking
sites, such as LinkedIn, MySpace and Facebook, which allow users to
articulate their professional and personal relationships by adding
colleagues, classmates and friends to their social networks.

How do users use social networks?
Recent research has shown that people tend to use social networking
sites to find job recommendations or dates~\cite{boyd04}, to learn
more about people they know offline~\cite{lampe06} or simply keep in
touch with a diverse group of offline friends~\cite{PewInternet,ambient-intimacy}.
This appears not to be the case with social media sites. In fact, due to their geographic
diversity and the practice of using pseudonyms,
few of a typical user's contacts on a social media site are
known to her offline. Instead, we believe that social media users
\emph{create} social networks to express their preferences and interests,
and use them to \emph{filter} the vast stream of new
submissions to find interesting content.
In fact, by exposing user
activity, social media sites allow users to leverage the opinions
and knowledge of others to solve their own information processing
problems, such as information filtering and document rating.

This paper examines how users of two popular social media
sites --- the social news aggregator Digg (\texttt{http://digg.com})
and the social photosharing site
Flickr (\texttt{http://flickr.com}) --- use social networks.
We claim that rather than
searching for interesting new content by keywords (e.g., tags) or subscribing to special
interests groups, users browse through the content created by
their contacts via the \emph{Friends interface}.
The social  network-driven Friends interface allows a user to easily
track her friends' activity: the new content they recently created
or liked. In this way social networks facilitate new ways of interacting
with information --- what we call \emph{social browsing}. Social
browsing is one of the most important usage modalities of social
media sites.

The paper is organized as follows. In \secref{sec:anatomy} we
describe in detail the functionality and features of Flickr and
Digg. In \secref{sec:data} we describe the datasets and data collection
methods. We analyze the data from the two sites in
\secref{sec:browsing} to present evidence for the claim that social
networks are used for information filtering. \secref{sec:related}
compares our work to existing research in social networks and other
fields, and \secref{sec:conclusion} presents our conclusions.

\section{Features of social media}
\label{sec:anatomy}
Social media sites share common features, no matter what the content
that they allow users to share. We illustrate the commonalities on
two popular sites: the social news aggregator Digg and the social
photosharing site Flickr. Although there are fundamental differences
between the two sites --- on Flickr, for example, users share
content (images) that they themselves create, while on Digg they
share content (news stories) that others create --- there are also
many similarities, which are also found on other social media sites.

The news aggregator Digg relies on users to submit
 and moderate news stories. When a story is submitted, it goes to the
upcoming stories queue, where it sits while users vote on it.
When a story gets enough positive votes, or \emph{diggs}, it is promoted to the front page.
The vast majority  of people who visit Digg daily, or subscribe to
its RSS feeds, read only the front page stories; hence, getting to
the front page greatly increases a story's visibility. Although
the exact promotion mechanism is kept secret and changes
periodically, it appears to take into account the number of votes
the story receives. Digg's  front page, therefore, emerges by
consensus of many independent evaluators.

\begin{figure}
\begin{tabular}{cc}
\includegraphics[height=3.4in]{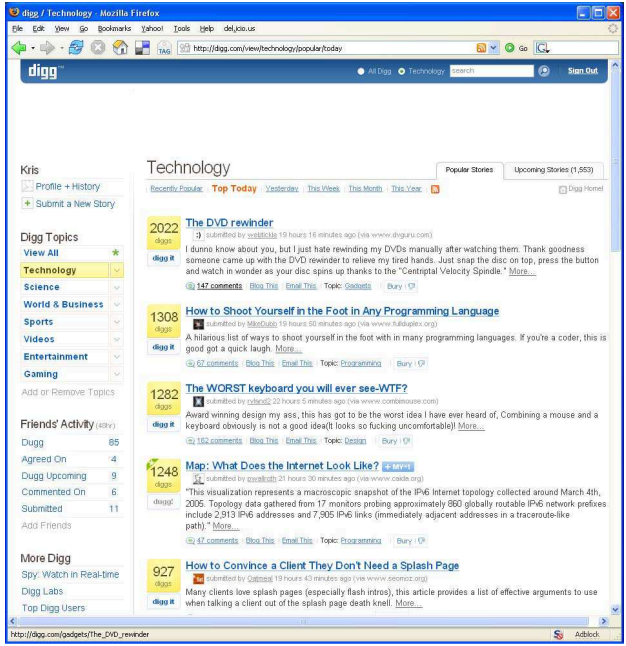} &
\includegraphics[height=3.4in]{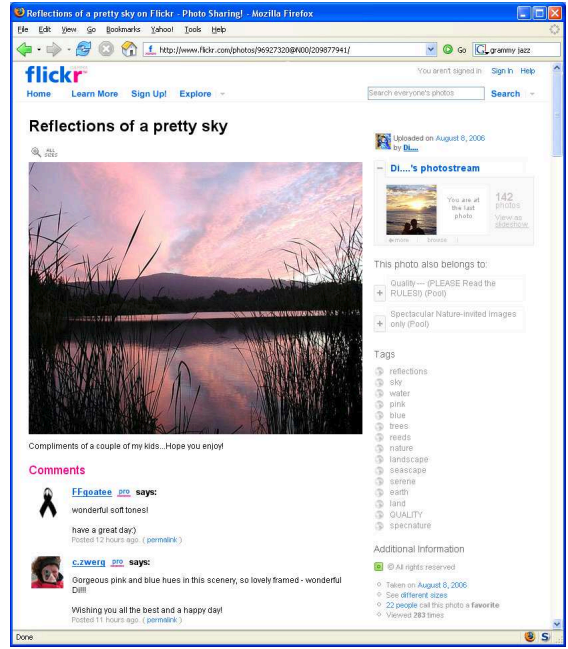} \\
(a) Digg front page & (b) Flickr photo page
\end{tabular}
\caption {Typical pages from (a) the social news aggregator Digg and
(b) the social photosharing site Flickr.
 } \label{fig:pages}
\end{figure}

A typical Digg front page is shown in \figref{fig:pages}(a).
The stories are displayed in reverse chronological order of being
promoted, 15 stories to the page, with the most recent story at the
top. The story's title is a link to the source, while clicking on
the number of diggs takes one to the page describing the story's
activity on Digg: the discussion around it, and the list of people who
voted on it.

Like other social media sites, Digg allows users to designate others
as \emph{friends} or \emph{contacts} and makes it easy
to track friends' activities.\footnote{Note
that the friend relationship is asymmetric. When user $A$ lists user
$B$ as a \emph{friend}, user $A$ is able to watch the activity of
$B$ but not vice versa. We call $A$ the \emph{reverse friend} of
$B$. If user $B$ also marks $A$ as a friend, then they are each other's
\emph{mutual friends}.} The left column of the front page in
\figref{fig:pages}(a), the so-called \emph{Friends interface}, summarizes
recent activity of user's friend: the number of stories they have
submitted, commented on or liked (dugg) within the past 48 hours.

Until February 2007 Digg ranked users
according to how many of the stories the user submitted were
promoted to the front page.  Clicking on the \emph{Top
Users} link allowed one to browse through the ranked list of users.
There is speculation that ranking increased competition,
leading some users to be more active in order to improve their
rank.  Digg discontinued making the list publicly
available, citing concerns that marketers were paying top users to
promote their products and services~\cite{WSJ}, although it is now
available through a third party~\footnote{http://www.efinke.com/digg/topusers.html}.

The photosharing site Flickr lets users upload,
manage and share photographs, participate in groups and discussions, etc.
Users manage photographs by assigning descriptive labels, called
\emph{tags}, to them.
A typical Flickr photo page, shown in \figref{fig:pages}(b), provides a variety
of information about the photo: who uploaded it and when, what
groups it has been submitted to, its tags, who commented on the
image and when, how many times it was viewed or marked as a
``\emph{favorite}''. A user page shows the latest photos she has
uploaded, the images she marked as her favorite, and
her profile, which includes a list of their contacts and
groups they belong to. A tag page shows either that user's images
tagged with a specific keyword, or all public images that have been
similarly tagged. A group page shows the photo pool, group
membership, popular tags, discussions and other information about
the group. Finally, the \emph{Explore} page has the calendar view which
shows the 500 most ``\emph{interesting}'' images uploaded on any day.
Like Digg, Flickr uses a secret formula that analyzes user activity
to identify most ``interesting'' images.
The \emph{Interestingness} algorithm is kept secret to prevent gaming
the system, but it takes into account ``where the
clickthroughs are coming from; who comments on it and when; who
marks it as a favorite; its tags and many more things which are
constantly
changing.''\footnote{http://flickr.com/explore/interesting/} In
addition to these browsing methods, Flickr also provides a map
interface that allows users to browse geotagged images.

Like Digg, Flickr allows users to designate others as contacts,
and offers an interface equivalent to Digg's \emph{Friends interface},
which enables users, with a single click, to see the latest images from their
contacts. Unlike Digg, Flickr supports different types contact
relationships, with different levels of privacy. A Flickr user can
mark another user as  a ``contact,'' ``friend,'' or ``family.'' Images
can be marked as ``private'' (visible to the user who uploaded the image only),
``friends'' and/or ``family'' only (visible to user's designated friends
and/or family respectively), or ``public,'' meaning that they
are visible to all of user's contacts through the Friends interface,
or to anyone else who navigates to the image page.

\section{Data and statistics}
\label{sec:data}

For our study, we tracked story and image activity
on Digg and Flickr by collecting data from the sites either through
specialized Web scrapers, or through an API provided by the site.
Because Digg did not provide an API until April 2007, we collected data
from it by scraping the site
with the help of Web wrappers, created using
tools provided by Fetch Technologies\footnote{http://fetch.com/}. We
extracted the following data:

\begin{description}
  \item[Digg-frontpage:]  a list of stories
from the first 14 pages of Digg. For each story, we extracted
submitter's name, story title, time submitted, number of votes and
comments the story received.

  \item[Digg-all:]  a list of stories
from the first 20 pages in the upcoming stories queue. For each
story, we extracted the submitter's name, story title, time
submitted, number of votes and comments the story received.

  \item[Top-users:]  information about the top 1020 of
the recently active users.
  For each user, we extracted the number of stories
  the user has submitted, commented and voted on; number of stories
promoted to the front page; users's rank; the list of
  friends, as well as reverse friends or ``people who
  have befriended this user.''
\end{description}

\emph{Digg-frontpage} and \emph{Digg-all} wrappers were executed
hourly over a period of a week in May  and in July 2006.

We used the Flickr API to download a variety of data for our study.
For the data not provided through the API (for example, the number
of views), we wrote specialized scrapers to extract this
information from the Web pages. Since scraping required a separate
HTTP request, this had an effect on the image statistics (e.g.,
number of views is incremented by every HTTP request). We corrected
for this effect in post-processing. We gathered the following data from Flickr:
\begin{description}

\item[Explore set:] consisted of the 500 ``most
interesting'' images (as chosen by Flickr's \emph{Interestingness}
algorithm) uploaded on July 10, 2006. We saved the image's rank on
the first day (the lower the rank, the more interesting the image).

\item[Apex set:] consisted of the 500 most recent images added to the
\source{Apex}
group\footnote{http://www.flickr.com/groups/apexgroup/}. This group
is one of ``the best of Flickr'' groups that are intended to to
showcase the best images and photographers.
Photographs can be added to the group only by invitation from
another group member.

\item[Random set:] contains 480 most recent of the images
uploaded to Flickr on July 10, 2006 around 4 pm Pacific Time.
Although we started with 500 images, some were made private or
deleted entirely from Flickr, leaving us with a smaller set.

\end{description}

For each image, we collected the name of the user who uploaded the
image (\emph{image owner}); the number of views and comments the image received; number of
times it was marked a ``favorite''; how many tags it had; the number
of groups it was submitted to. We also extracted the names of users
who commented on or favorited the image. We also tracked the number of
views, comments and favorites received by images in the three
datasets hourly over the period of
eight days starting July 10, 2006.

In addition to image statistics, we extracted data about Flickr's
social networks. While the site shows a user's list of contacts, one
cannot easily get the list of user's reverse contacts, i.e., other
users who list the particular user as a contact. This is important
information, since it shows how many people have access to the
user's photo stream. In order to reconstruct reverse contacts, we
crawled Flickr's network of contacts. We limited the crawl to depth
two due to the explosive growth of the network. Starting with about
1,100 unique users from our three datasets, we downloaded these
users' contacts, and their contacts' contacts. This gave us a
network with over 55,000 unique users and 5,000,000 connections. The
resulting social network is not complete, but it allows us to
estimate the number of reverse contacts a user has.

\subsection{Content activity}
\begin{figure}[tbh]
\begin{tabular}{cc}
  \includegraphics[height= 2.1in]{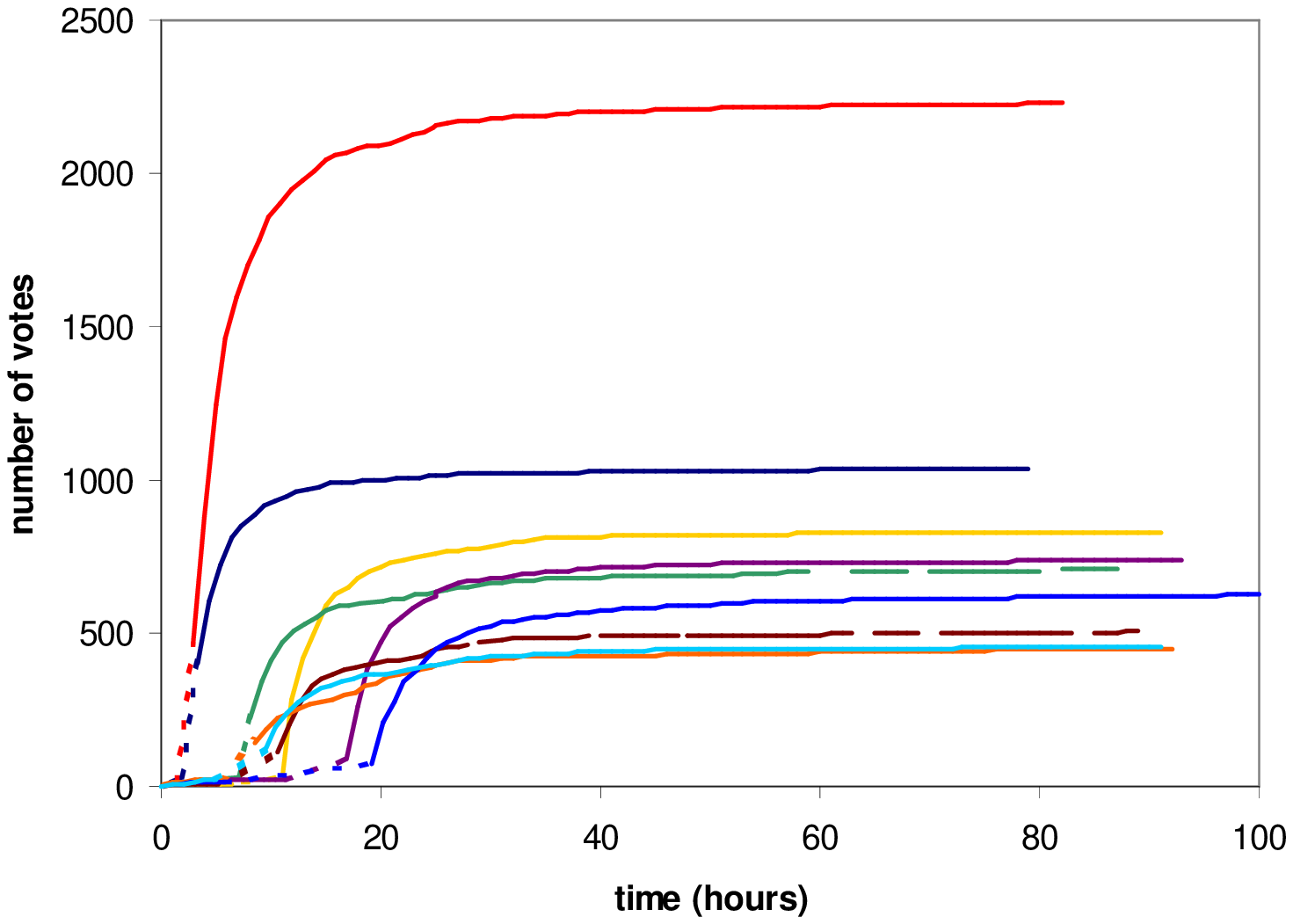} &
  \includegraphics[height= 2.1in]{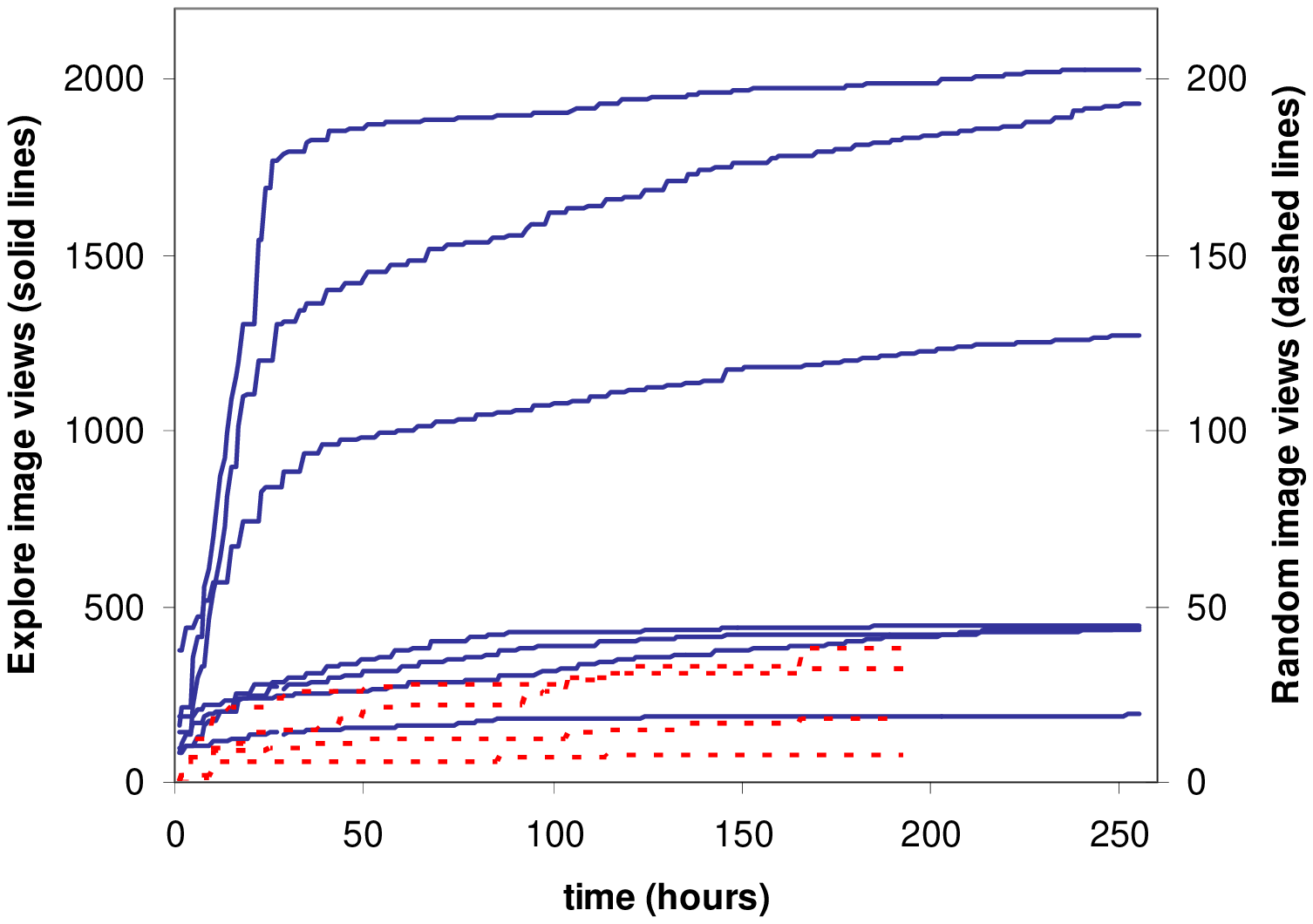} \\
  (a) votes on Digg & (b) views on Flickr
  \end{tabular}
\caption{ (a) Dynamics of votes received by select stories on Digg over a period
of four days. Dashes indicate story's transition to the front page.
(b) Cumulative number of times images in the \source{Explore} set (solid lines) and
  \source{Random} set (dashed lines) were viewed over the time of the tracking period
} \label{fig:dynamics}
\end{figure}

We identified stories that were submitted to Digg over the course of
approximately one day and followed them over
several days. Of the $2858$ stories submitted by $1570$
distinct users over this time period, only 98 stories by 60 users
made it to the front page. \figref{fig:dynamics}(a) shows evolution of
the number of votes received by a few randomly chosen stories from the set.
The basic dynamics of all
stories appears the same. While in the upcoming queue, a story
accrues votes at some slow rate. Once it is promoted to the front page
(indicated by dashes in \figref{fig:dynamics}(a)),
it accumulates votes at a much faster rate. As the story ages,
accumulation of new votes slows down~\cite{Wu07}, and the story's rating
saturates at some value, indicative of how interesting the story is to the general Digg community.

\comment{
It is worth noting that the top-ranked users are not submitting the most interesting stories
(that get the most votes). Slightly more than half of the stories our data set came
from 14 top-ranked users (rank$<25$) and 48 stories came from 45
low-ranked users. The average ``interestingness'' of the stories
submitted by the top-ranked users is $600$, almost half the average
``interestingness'' of the stories submitted by low-ranked users. A
second observation is that top-ranked users are responsible for
multiple front page stories. A look at the statistics about top
users provided by Digg shows that this is generally the case: of the
more than 15,000 front page stories submitted by the top 1020
users, the top $3\%$ of the users are responsible for $35\%$ of the
stories.
}

We tracked the number of
views, comments and favorites received by images in the three Flickr
datasets hourly over the period of
eight days starting July 10, 2006.

The number of views received by the newly uploaded images on Flickr followed a pattern similar to
that of votes received by stories on Digg.
\figref{fig:dynamics}(b) shows the number of times randomly chosen
images from the \source{Explore} and \source{Random} sets were
viewed (the number of views received by \source{Apex} images, some of them
months old, did not change much over the course of the tracking period).
The curves are jagged because Flickr updates
the counts of views every two hours. Images generally received most
of their views within the first two days, after which they were
viewed much less frequently, except for some \source{Explore} images.

\begin{figure}[tbh]
\center
  \includegraphics[width=4.0in]{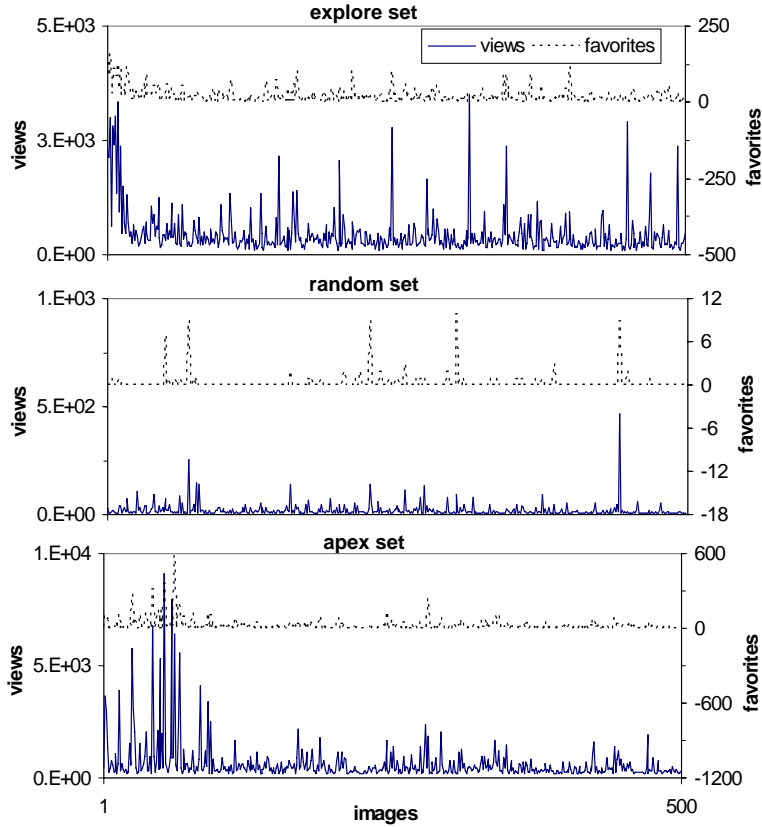}

  \caption{The number of times images in the \source{Explore}, \source{Apex} and \source{Random} sets were viewed
  and favorited by the end of the tracking
  period. Images in the \source{Explore} set are sorted by their rank, while \source{Apex} and \source{Random}
  images are shown in their chronological
  order of being added to the group or uploaded to Flickr respectively.}\label{fig:views}
\end{figure}

The top-rated \source{Explore} images showed
the ``Explore effect,'' a dramatic rise in the number of views
received by images featured on the \emph{Explore} page, Flickr's version of the front page.
The ``Explore effect'' is pronounced in \figref{fig:views},
which shows the total number of times the images in each set were
viewed over the course of eight days. While images in the
\source{Random} set received on average just 20 views, the
\source{Explore} images (ordered by their \emph{Explore} rank) received on average 450 views.
\source{Apex} images show cumulatively more views because they are much older,
although there was no significant increase in the number of views
over the course of the tracking period.
The top 20 \source{Explore} images show the biggest overall gain in
views. This is probably caused by the following factors: (a)
the 10 top ranked images can be posted to the special
\emph{Interestingness| Must be in Top
10} group,\footnote{http://www-us.flickr.com/groups/interestingness/}
(b) people who browse Flickr through the calendar interface
probably scan the first two pages of images (10 images on each page)
without paging further, or most
likely because (c) the popular and prominent \emph{Explore} page
features one of the top
20 images from the previous and current days picked at random.

The number of times an image has been marked as a favorite (dotted
lines in \figref{fig:views}) generally follows the number of views.
Marking an image as a favorite is Flickr's
analog of voting on the image, although number of favorites is only
part of the formula used to select images for the \emph{Explore} page.

\subsection{Social networks}
\label{sec:socnetsdata}
\begin{figure}[tbh]
\begin{tabular}{cc}
  \includegraphics[height=2.4in]{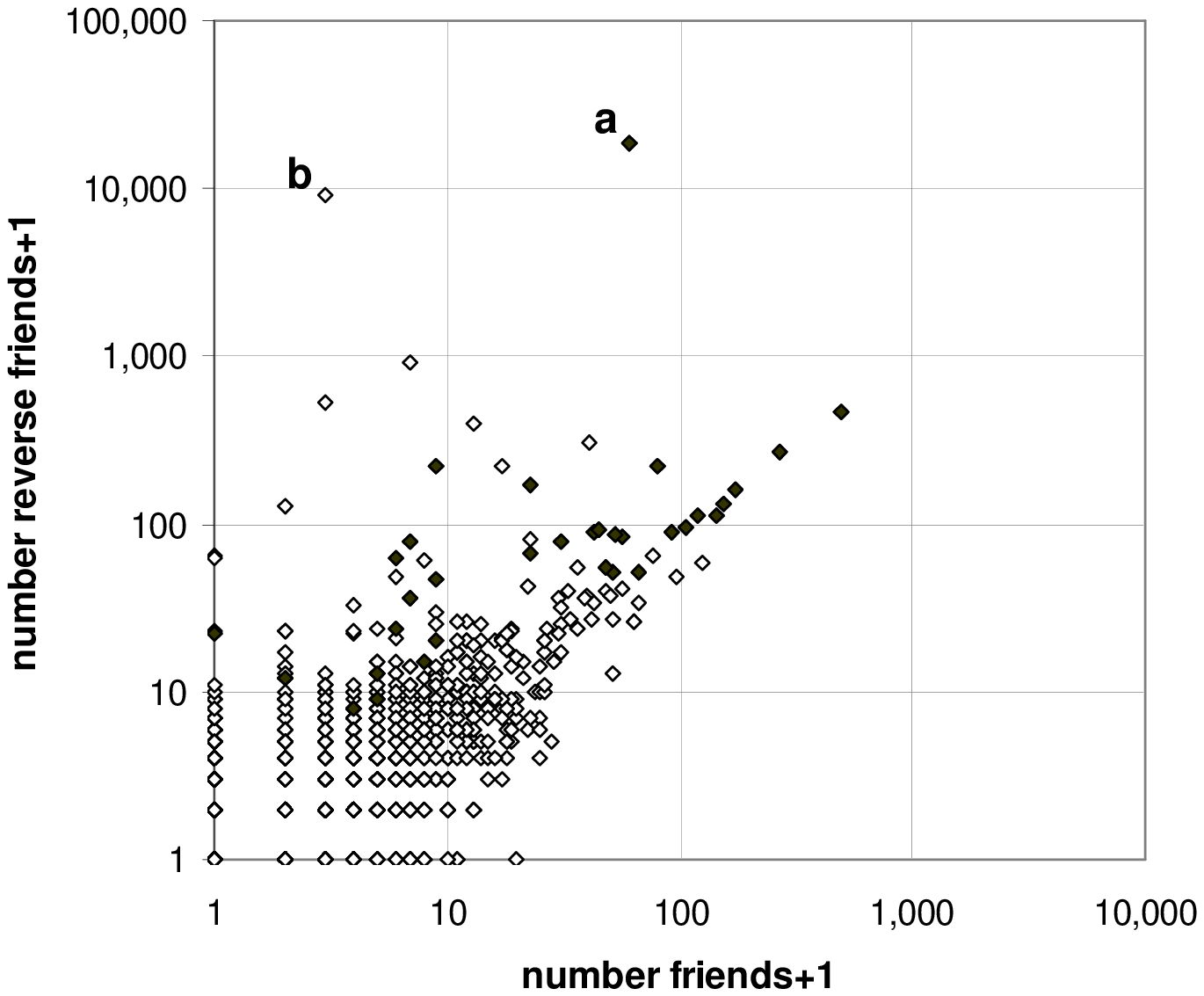} &
  \includegraphics[height=2.4in]{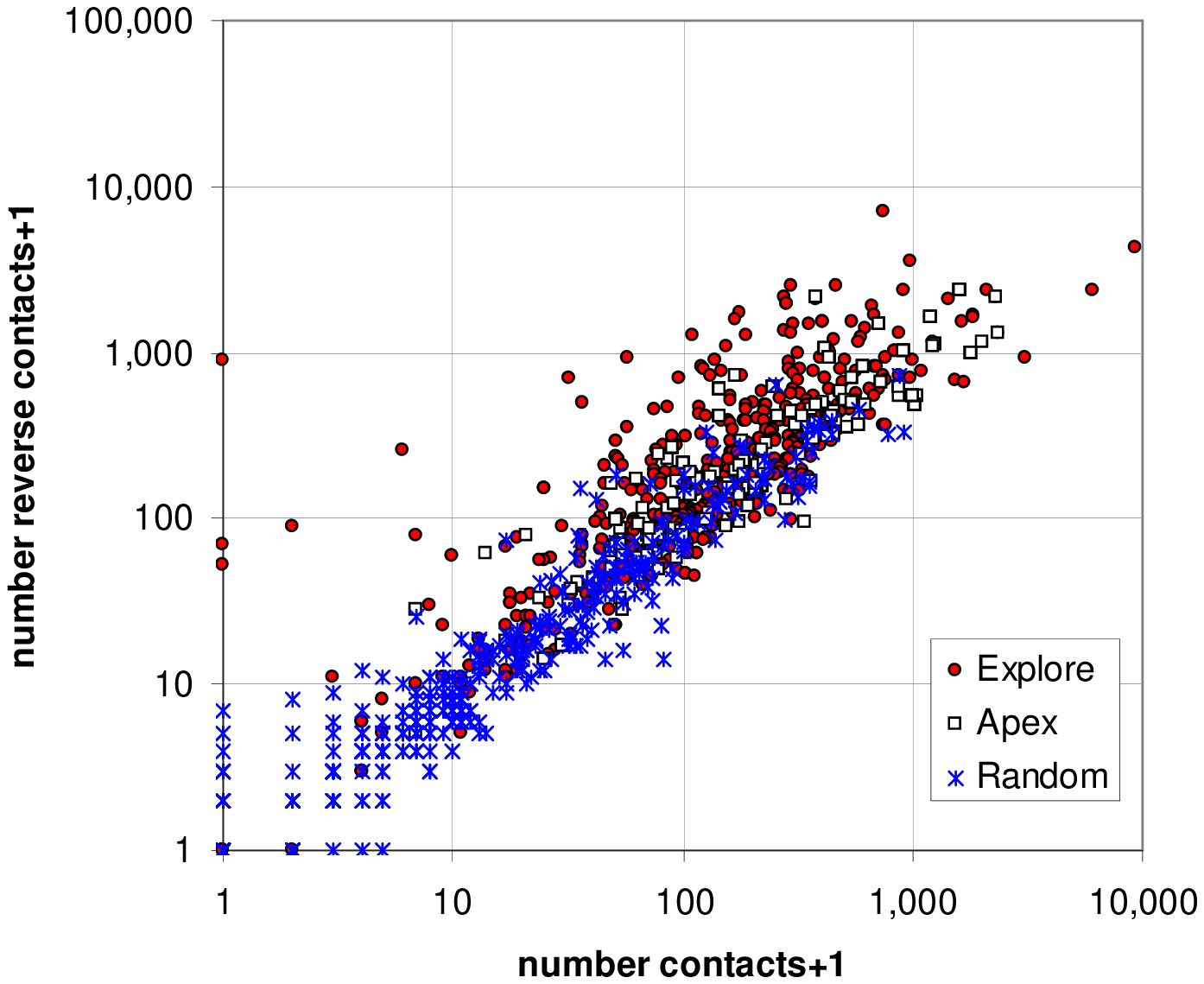}\\
  (a) Digg & (b) Flickr \\
\end{tabular}
  \caption{Scatter plot of the number of friends (contacts) vs reverse friends (contacts) for (a) the top 1020 Digg users
 and  (b) 1100 Flickr users from the \source{Apex}, \source{Explore} and \source{Random} datasets.
story}\label{fig:scatterplot}
\end{figure}

Do users take advantage of the social networking feature of social
media sites?
\figref{fig:scatterplot}(a) shows the scatter plot of
the number of friends vs reverse friends of the top 1020 Digg users
as of May 2006. Black symbols correspond to the top 33 users. For
the most part, users appear to add friends to their social networks,
with the top users having bigger social networks.
Two of the biggest celebrities (watched by most people) are users marked $a$ and $b$ on
\figref{fig:scatterplot}(a), corresponding $kevinrose$ and
$diggnation$, respectively, one of the founders of Digg and a
podcast of the popular Digg stories.

\figref{fig:scatterplot}(b) shows a similar scatter plot for the approximately 1,100 unique Flickr
users who uploaded images in our three datasets.
The number of reverse
contacts is not directly available and was estimated by crawling the
contacts network as explained above. Generally,
users in all three datasets had contacts and were listed as contacts
(reverse contacts) by other users, with \source{Explore} and
\source{Apex} users being better connected than \source{Random}
users. The points are scattered around the diagonal, indicating
equal numbers of contacts and reverse contacts (possibly indicating
mutual contact relationships), although \source{Apex}, and especially
\source{Explore}, users had greater numbers of reverse
contacts. Note also that Flickr users appear to be more active in creating social networks compared to Digg users.
\comment{\footnote{Interestingly, four of the images in the
\source{Explore} set came from users with no reverse contacts, and
two of these were not shared with any groups. Both of these images
were about pandas, and were tagged with ``panda.'' This shows either
that panda aficionados on Flickr are active and do use tags to
search for new images of pandas, or people behind Interestingness
algorithm chose pandas as the featured animal of the month.}}

\section{Social networks and information filtering}
\label{sec:browsing}
What do people do with the social networks they create?
We claim that on social media sites, social networks are used for \emph{information
filtering} ---
to select from the vast stream of new submissions the
content that the user will likely find interesting. The filtering is
accomplished via the mechanism of social browsing, which simply
means using the Friends interface to browse the social media site.
We claim that social browsing one of the most important browsing
modalities in social media. This has implications on
other aspects of social media, e.g., how content and users are ranked, and how content is selected to be
featured on the front page.

Below we present evidence to support our claims.
First, we present indirect evidence for social browsing on Digg by showing that
user's success is correlated with his social network size.
We then present additional evidence that social browsing (through the Friends
interface) is used for information filtering, by showing that users
tend to like the stories their friends like.
In \secref{sec:socialbrowsingflickr} we analyze data from the
\source{Random}, \source{Apex} and \source{Explore} image sets to show that social
browsing explains much of the activity generated by new images on Flickr.

\subsection{Social browsing on Digg}
\label{sec:socialbrowsingdigg}

\begin{figure}[tbh]
\center
\begin{tabular}{cc}
  \includegraphics[height=2.0in]{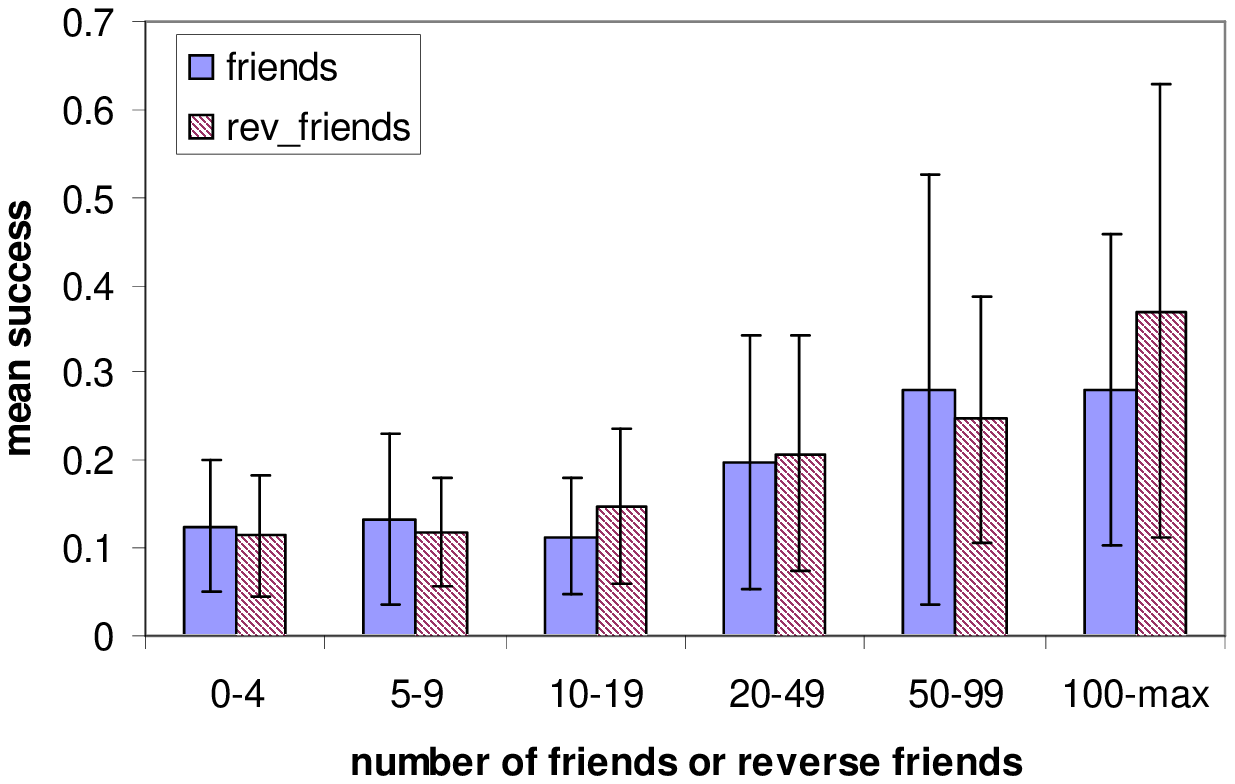} &
    \includegraphics[height=2.0in]{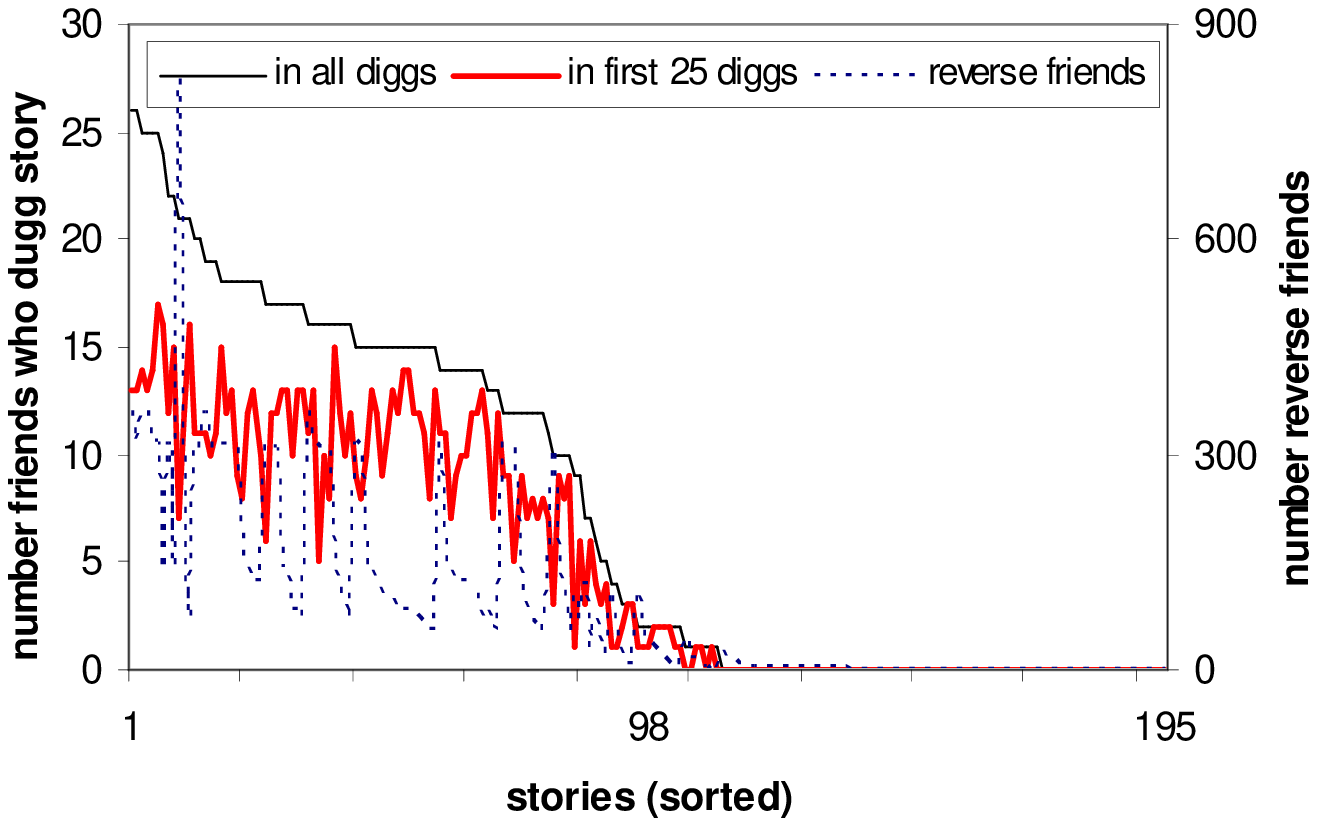} \\
 (a) & (b)
 \end{tabular}
  \caption{(a) Strength of the linear correlation coefficient between user's success rate and the number
  of friends and reverse friends he has. (b) Number of voters who are also among
  the reverse friends of the user who submitted the story}\label{fig:digg-correlation}
\end{figure}

 A user's success
rate is defined as the fraction of the stories the user submitted
that have been promoted to the front page. We use the statistics
about the activities of the top 1020 users to show that users with
bigger social networks are more successful at getting their stories
promoted to the front page. We only
include users who have submitted 50 or more stories (total of 514
users). The correlation between users's mean success rate and the size of their social
network is shown in \figref{fig:digg-correlation}. Data was binned to improve statistics. Despite large error
bars, there is a significant correlation between users's success rate and the size
of their social network, more importantly, the number of
reverse friends they have. In \cite{Lerman07ic} we constructed a
phenomenological model of the dynamics of votes received by stories
on Digg. We showed that users with bigger social networks could have
less interesting stories promoted to the front page. The high degree
of correlation between promotion success rate and submitter's social
network size is in line with those findings.

In the sections below we present a more direct evidence that the Friends interface
is used to find new interesting stories. We show this by analyzing two
sub-claims: (a) \emph{users digg stories their friends submit}, and
(b) \emph{users digg stories their friends digg}.
By ``digging'' the story, we mean that users like the story and vote on it.

\subsubsection{Users digg stories their friends submit} We scraped Digg to collect
data about 195 front page stories, including a list of the first 216 users to vote
on each story ($15,742$ distinct users in total). The name of the submitter
is first on the list.

We can compare this list, or any portion
of it, with the list of the reverse friends of the submitter.
The dashed line in \figref{fig:digg-correlation}(b) shows the size of the social
network (number of reverse friends) of the submitter. More than half
of the stories (99) were submitted by users with more than 20
reverse friends, and the rest by poorly connected users.\footnote{These users
have rank $>1020$ and were not listed as friends of any of the 1020
users in our dataset. It is possible, though unlikely, that they
have reverse friends.}
The thin line  shows the number of voters
who are also among the reverse friends of the submitter.
All but two of the stories (submitted by users with
47 and 28 reverse friends) were dugg by submitter's reverse friends.

We use simple combinatorics \cite{Papoulis} to compute the probability that
$k$ of submitter's reverse friends could have voted on the story purely by
chance. The probability that after picking $n=215$ users randomly
from a pool of $N=15,742$ you end up with $k$ that came from a group
of size $K$ is $ P(k,n)={n\choose k} (p)^k (1-p)^{n-k}$, where
$p=K/N$. Using this formula, the probability (averaged over stories
dugg by at least one friend) that the observed numbers of reverse friends
voted on the story by chance is $P=0.005$, making it highly
unlikely.\footnote{If we include in the average the two stories that
were not dugg by any of the submitter's friends, we end up with a
higher, but still significant P=0.023.} Moreover, users digg stories
submitted by their friends very quickly. The heavy red line in
\figref{fig:digg-correlation}(b) shows the number of reverse friends who
were among the first 25 voters. The probability that these numbers
could have been observed by chance is even less --- $P=0.003$. We
conclude that users digg --- or tend to like --- the stories their friends submit.
As a side effect, by enabling users to quickly digg stories
submitted by friends, social networks play an important role in
promoting stories to the front page.

\subsubsection{Users digg stories their friends digg}  Do social networks also help
users discover interesting stories that were submitted by poorly-connected
users? Digg's Friends interface allows users to see the stories their friends have liked (dugg).
As well-connected users digg stories submitted by users who have few or no reverse friends, are others
within his or her social network more likely to read them?

\begin{figure}
\begin{tabular}{cc}
  \includegraphics[height=2.in]{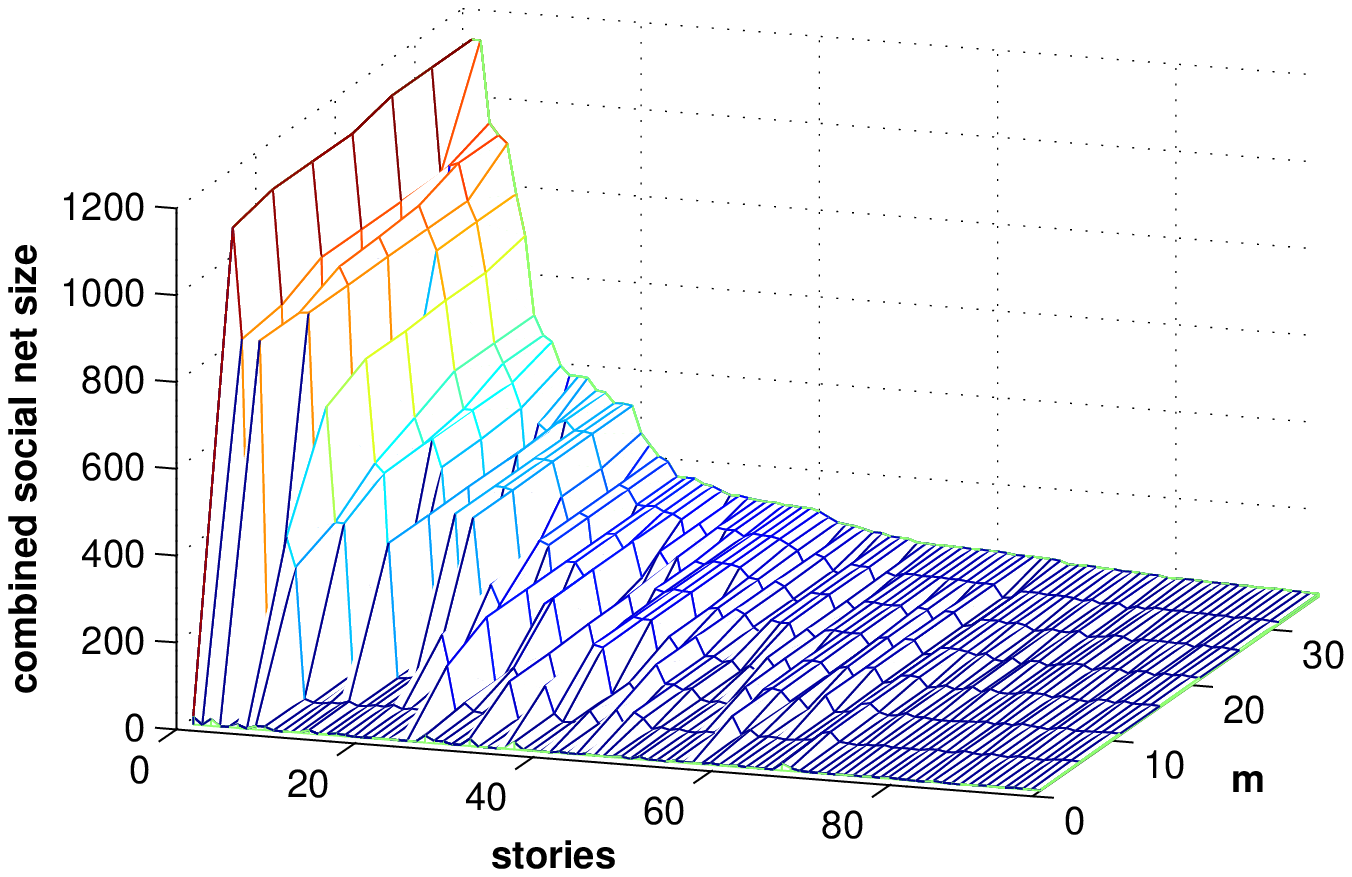}  &
  \includegraphics[height=2.in]{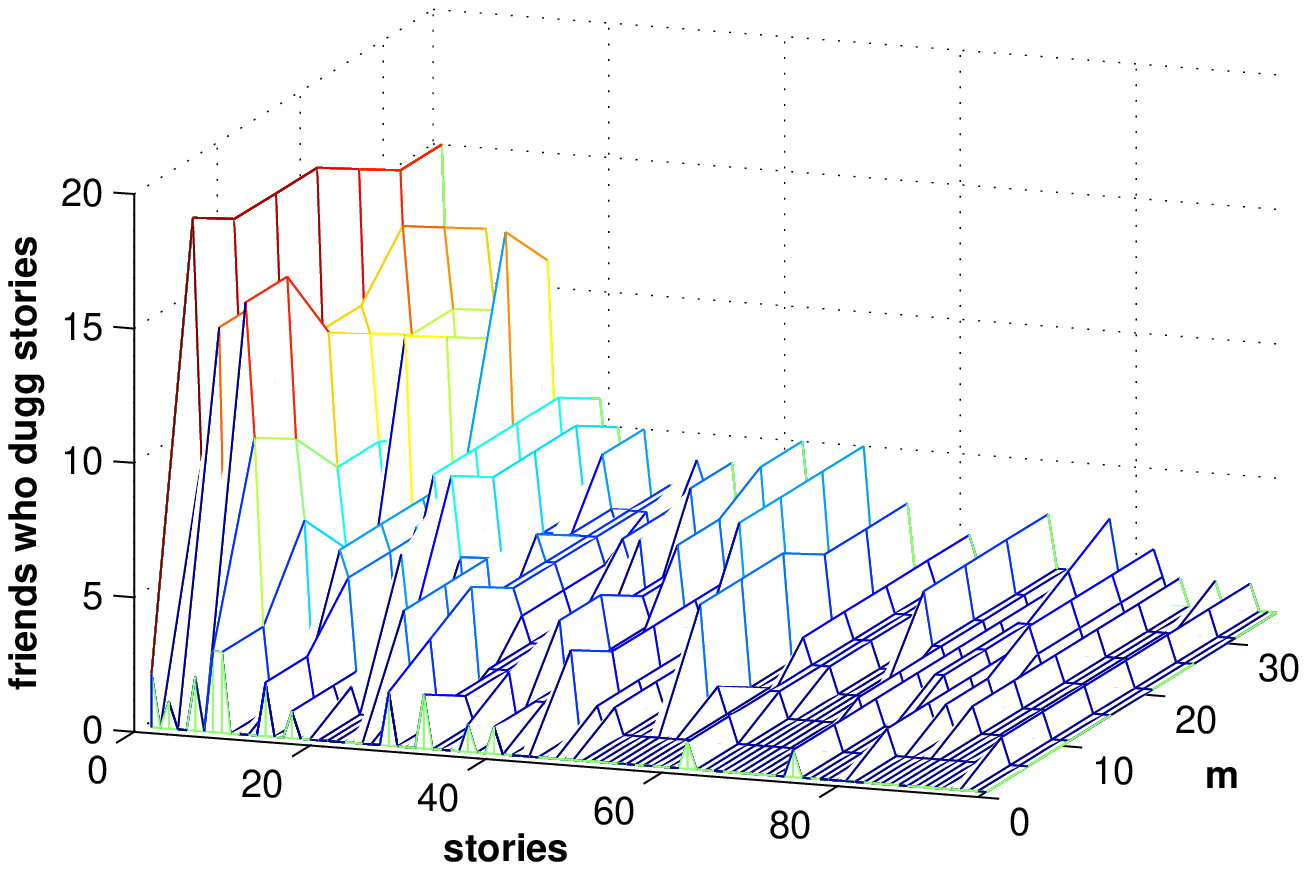}\\
  (a) & (b) \\
\end{tabular}
  \caption{(a) Number of reverse friends of the first $m$ voters for the stories submitted by poorly-connected users.
  (b) Number of friends of the first $m$ voters who also voted on the stories.}\label{fig:history-unknown}
\end{figure}

\begin{table*}
  \centering
\begin{tabular}{|ll|c|c|c|c|c|c|}
  \hline
  & \textbf{diggers} & \textbf{m=1} & \textbf{m=6} & \textbf{m=16} & \textbf{m=26} & \textbf{m=36} 
  \\ \hline
(a) & visible to friends & 34 & 75  & 94 & 96  & 96  \\
 (b) & dugg by friends & 10 & 23 & 37 & 46 & 49 \\
  (c) & probability & 0.005 & 0.028 & 0.060 & 0.077 & 0.090 
  \\ \hline
\end{tabular}

  \caption{Number of stories posted by poorly-connected users that were (a) made visible to others by
  digging activities of well-connected users, (b) dugg by friends of the first $m$ diggers within the next 25 diggs,
  and for the stories that were dugg by friends, (c) the average probability that the observed numbers of friends
  dugg the story by chance }\label{tbl:history-unknown}
\end{table*}

\figref{fig:history-unknown} shows how the activity of
well-connected users affected the 96 stories submitted by poorly-connected users, those with fewer than 20 reverse friends.
$m=1$ corresponds to the user who submitted the story, while $m=6$
corresponds to the story's submitter and the first five users to
digg it. \figref{fig:history-unknown}(a) shows how the combined
social network (number of reverse friends) of the first $m$ diggers grows as the story
receives votes. \figref{fig:history-unknown}(b) shows how many of
the following 25 votes come from users within the combined social
network of the first $m$ voters.

 At the time of submission ($m=1$), only 34 of the 96
stories were visible to others within the submitter's social
network and ten of these were dugg by submitter's reverse friends within the first 25 votes.
After fifteen more users have voted, almost all stories are now visible through
the Friends interface.
\tabref{tbl:history-unknown} summarizes the observations
and presents the probability that the observed numbers of reverse friends
voted on the story purely by chance. The probabilities for $m=26$ through $m=36$ are
above the $0.05$ significance level, possibly reflecting story's
increased visibility on the
front page. Although the effect is not quite as dramatic as one in
the previous section, we believe that the data indicates that users do
use the ``see the stories my friends have dugg'' portion of the Friends
interface to find new interesting stories.

\subsection{Social browsing on Flickr}
\label{sec:socialbrowsingflickr}
How do users find new images on Flickr? Do they
find them through groups, or by searching by tags? Do they find them
through the Explore page? Or by
by browsing through the photo streams of their contacts? We believe
that social browsing explains much of the activity generated by new images on Flickr.
We present a detailed study of the images from the
\source{Random}, \source{Apex} and \source{Explore} sets that help
answer these questions.

\subsubsection{Pools and tags}
\label{sec:pools}

\begin{figure*}[tbhp]
\center
 \begin{tabular}{cc}
\includegraphics[width=3.2in]{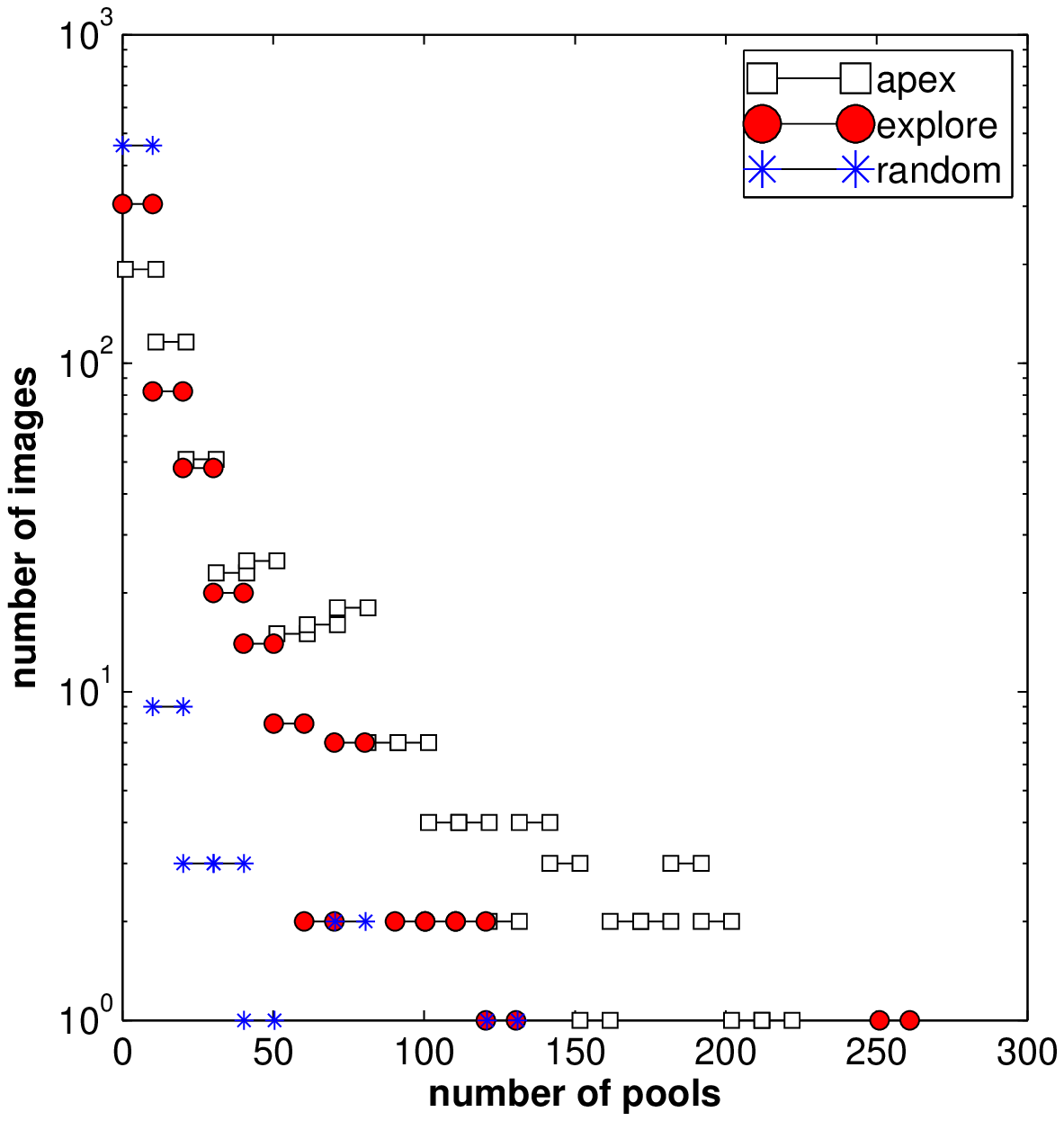} &

  \includegraphics[width=3.2in]{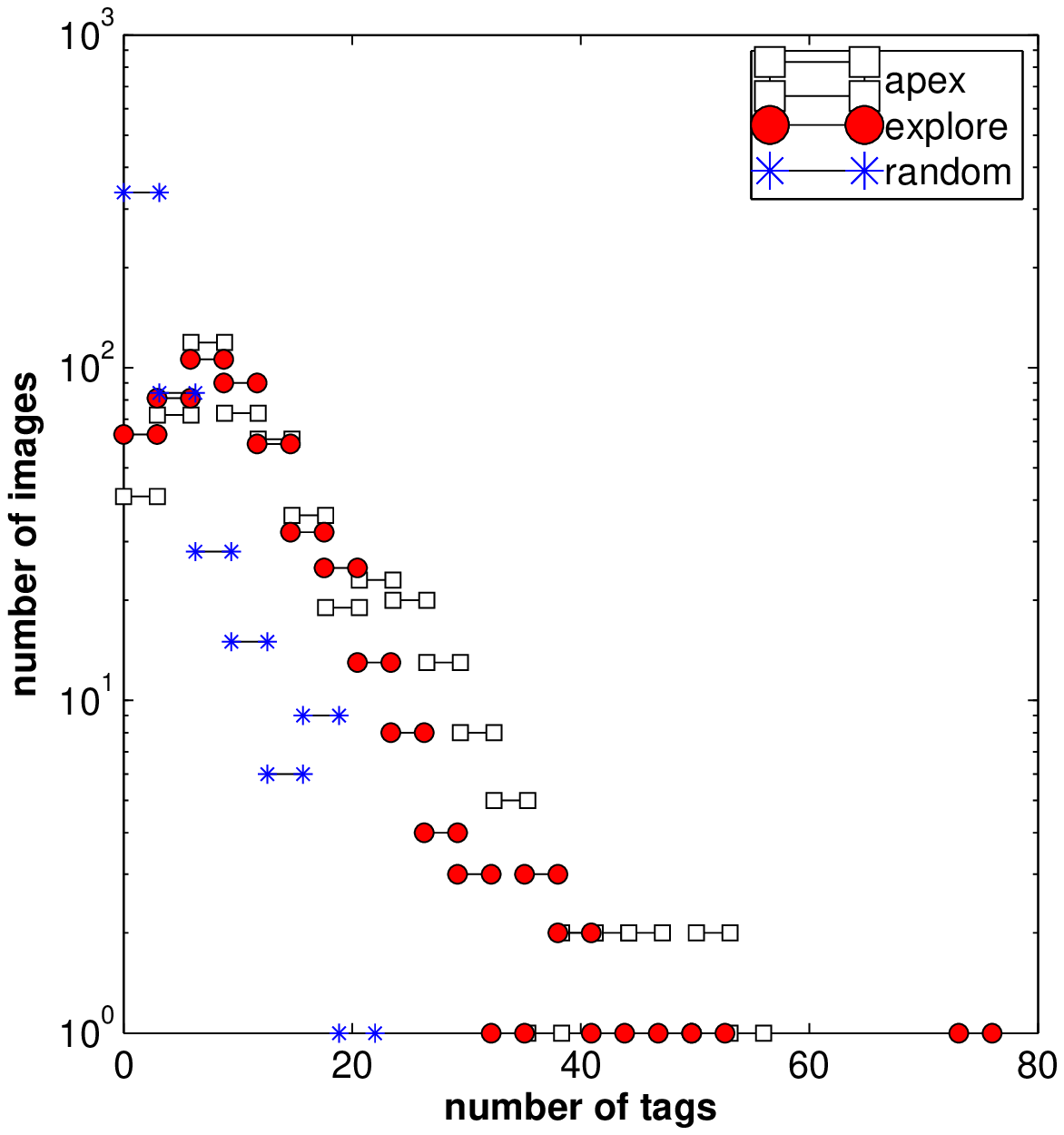} \\
  (a) & (b)

\end{tabular}
  \caption{Histogram of (a) the number of pools to which images from each set were submitted and
  (b) the number of tags assigned to the images}\label{fig:histogram}
\end{figure*}

When users upload images to Flickr, they have an option to share
them with different groups, each with its own image pool. A large
number of special interest groups already exist on Flickr, on a wide
variety of topics --- everything from Macro Flower Photography to
one dedicated to the color orange --- with new ones added daily.
There is often a substantial overlap between group interests (there
are more than a dozen groups dedicated to flowers alone), which
results in images being posted to multiple groups.
\figref{fig:histogram}(a) shows the distribution of the number of
pools to which images in the \source{Explore}, \source{Apex} and
\source{Random} sets have been posted. Although a typical user
(\source{Random} set) does not share images with any groups, some
users submit images to a surprisingly large number of groups ---
several users in the \source{Explore} and \source{Apex} sets have
submitted their images to over 100, and on a few occasions over 200,
groups.

Flickr also allows users to tag their images with descriptive
keywords. Tagging is advocated by Flickr as a way to improve search
of the user's own, as well as other people's, images.
\figref{fig:histogram}(b) shows patterns in tagging usage across
different data sets. Although very few \source{Random} users tag
their images, \source{Explore} and \source{Apex} users do tend to
use many tags, sometimes as many as 70. Interestingly, there seems
to exist a preferred number of tags --- around ten --- for images in
the \source{Explore} and \source{Apex} sets.

In both their tagging activity, as well as in submitting images to
groups, \source{Explore} and \source{Apex} users are very similar to
each other and different from \source{Random} users. There is
considerable effort involved in sharing an image with a group,
suggesting that social aspects of Flickr, such as sharing images
with other users through groups and increasing the visibility of an
image is very important to users, possibly more than being able to
easily find them with tags.

\subsubsection{Social networks and views}
\label{sec:socnetviews}

\begin{figure}[tbh]
 \begin{tabular}{cc}
  \includegraphics[width=2.8in]{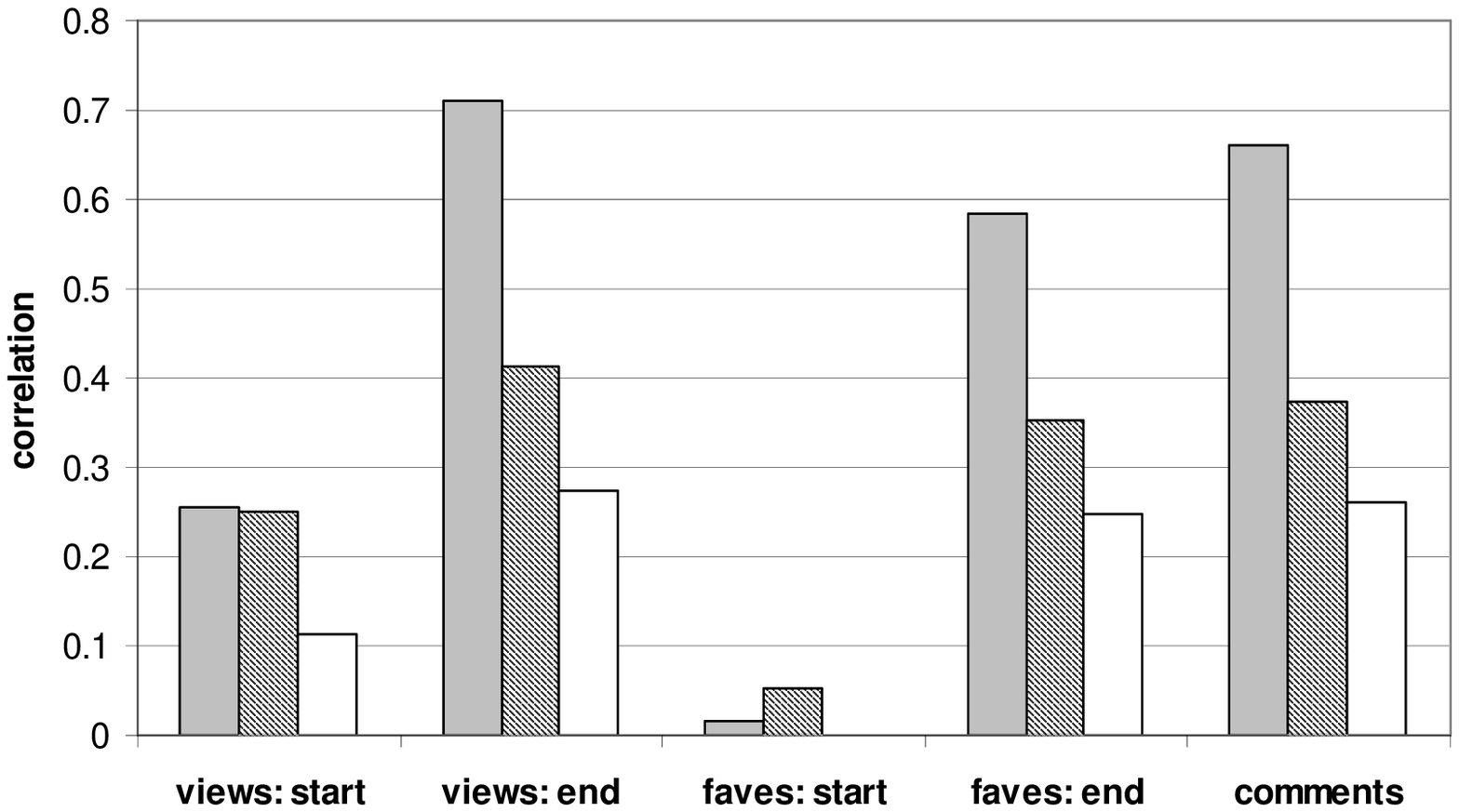} &
  \includegraphics[width=2.8in]{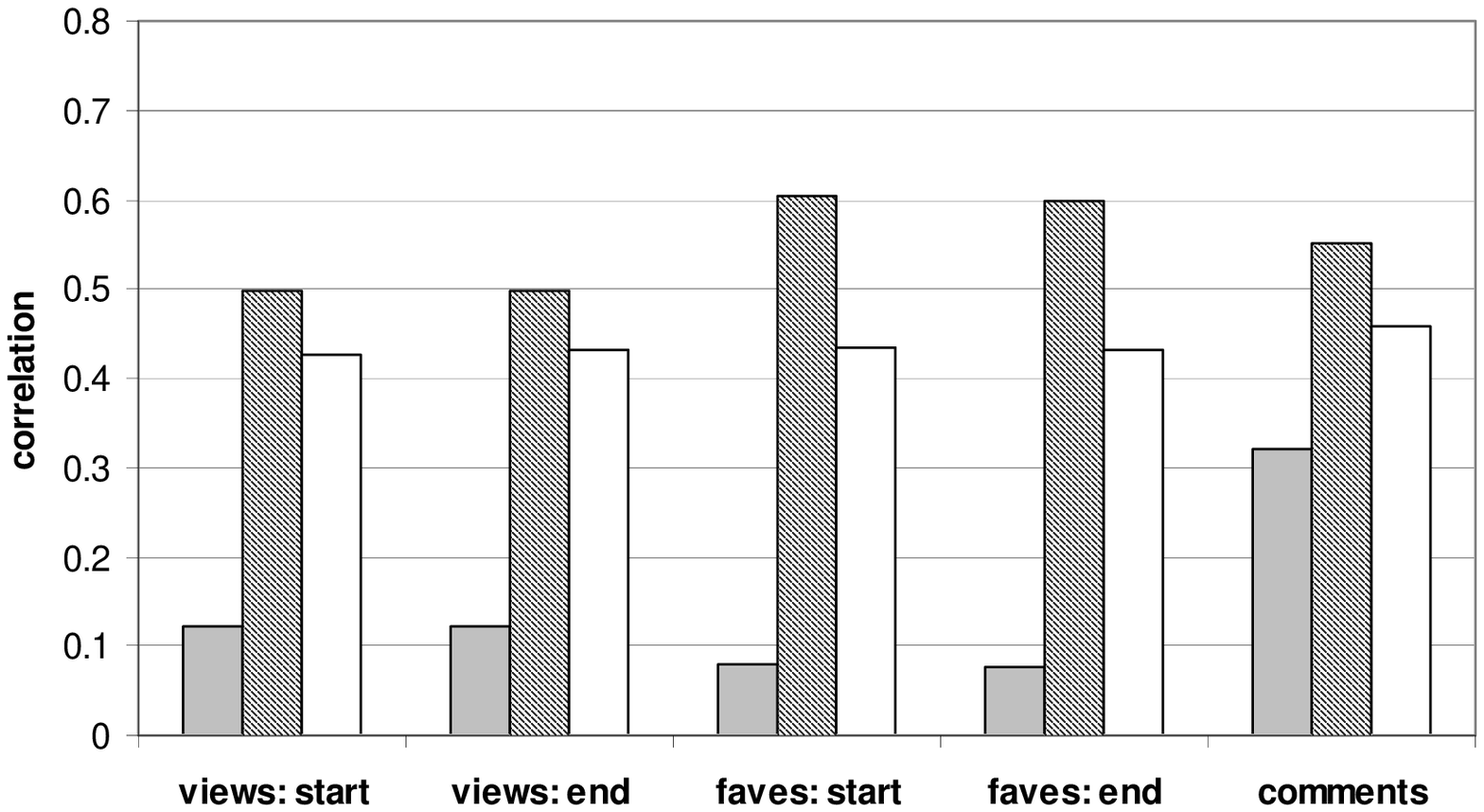} \\
  Random set  &  Apex set \\
  &\includegraphics[width=2.8in]{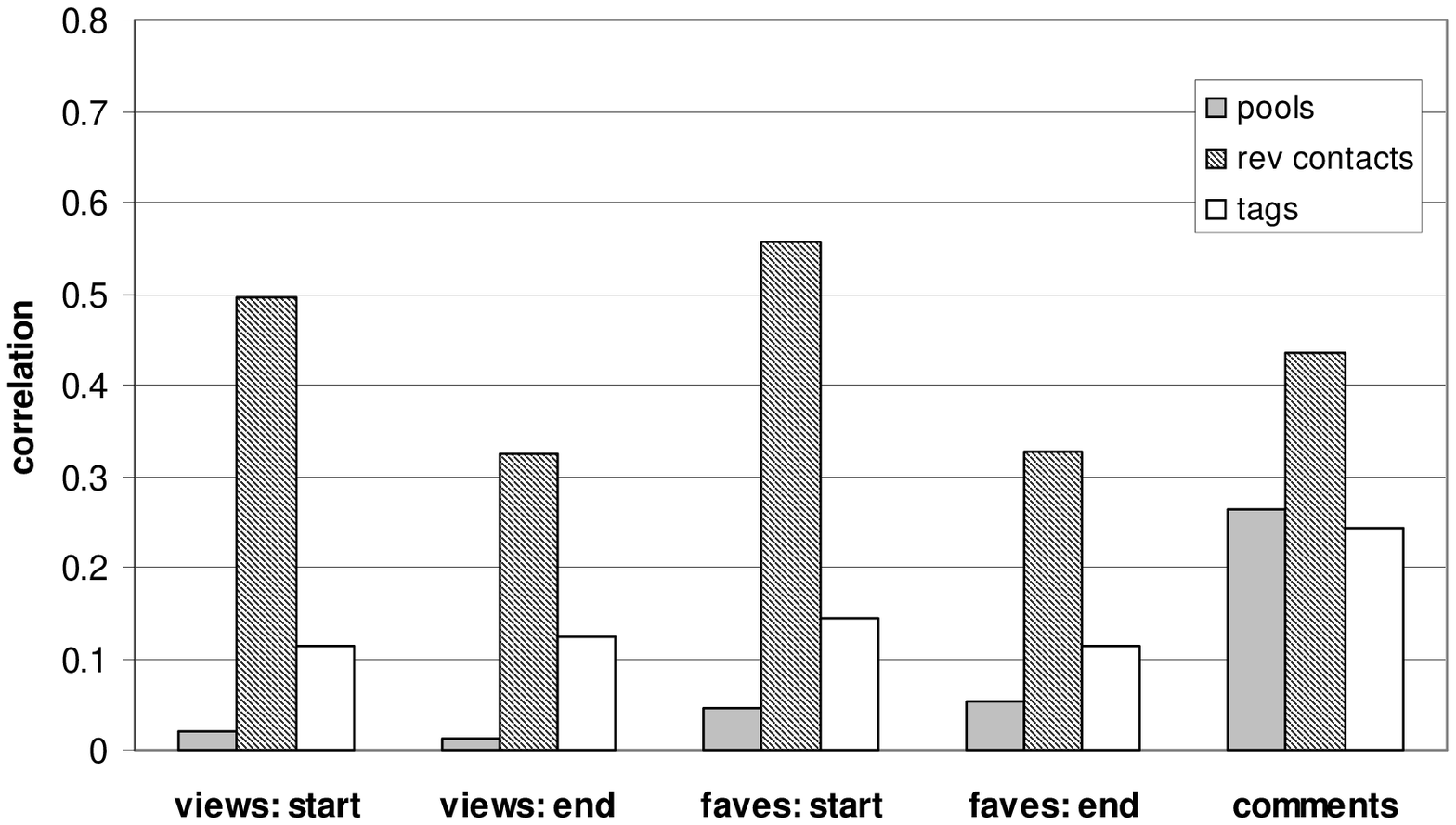} \\
  & Explore set \\
\end{tabular}
  \caption{Strength of the correlation between image statistics (number of views and favorites at the beginning and end of the tracking
  period, number of comments ) and image features
  (the number tags is has, pools it was submitted to and the size of the photographer's social network)
  for images in the three datasets. }
  \label{fig:flickr-correlation}
\end{figure}

We have shown (in \secref{sec:socnetsdata}) that users
generally take advantage of the social features of Flickr, adding others
as friends or contacts to their personal social networks. Flickr
gives users instant access to the latest images uploaded by their
contacts through the Friends interface (called ``Contacts'' on Flickr).
We claim that this inetrface is one of the more important browsing modalities on Flickr.
Unfortunately, Flickr does not provide a record of users who viewed an image.
Instead, we establish this link indirectly by showing a correlation
between the number of views generated by an image and the number of
reverse contacts the image's owner (user who submitted the image) has.
\figref{fig:flickr-correlation} shows the strength of the correlation
between image statistics and features, such as the number of
contacts and reverse contacts the image owner has,
the number of pools to which the image was submitted, and the number
of tags it was annotated with.\footnote{All the correlations with
correlation coefficient $C_r>0.1$ are statistically significant at
$0.05$ significance level. } The image statistics are: (1) the
number of views the image received and (2) the number of times it
was marked as favorite at the beginning and end of the tracking period and
(3) the number of comments it received.

\source{Apex} and the \source{Explore} sets show similar correlation
values at the start of the tracking period, where the number of
views, comments and the number of times the image was favorited
correlates strongly (or at least moderately) with the number of
reverse contacts the user has. At the end of the tracking period,
however, the number of views, favorites and comments for the images
in the \source{Explore} set is less strongly correlated with the
size of the user's social network. This could be explained by the greater
public exposure images receive through the Explore page. Groups seem
not to play a significant role in generating new views, favorites or
comments for these images. Tags appear to be uncorrelated with
image activity in the \source{Explore} set, but somewhat correlated
in the \source{Apex} set. This could be explained by users clicking
on the ``apex'' tag (that all \source{Apex} photos are required to
have) to discover new photos in that pool.

The data presented above shows that, at least until the image gets
to the Explore page, the number of views (favorites and
comments) that images produced by good photographers receive correlates
most strongly with the number of reverse contacts the image owner
has. This is best explained by social browsing, which predicts that
the more reverse contacts a user has, the more likely his or her
images are to generate views.\footnote{Flickr claims that its  \emph{Interestingness} algorithm takes
other factors besides views into account to reduce the presence of
popular photographers on Explore. Rather than using a complicated
formula mentioned earlier in this paper, they could simply compute
whether an image received a greater than expected number of views,
favorites and comments. This simple heuristic could help identify truly
exceptional images.} Views gathered by \source{Random}
images correlate most strongly to the number of pools the image was
submitted to, and only moderately to the number of reverse contacts.
Since users in the \source{Random} sets have smaller social
networks, they get more exposure by posting images to groups.


\begin{figure}[tbh]
 \begin{tabular}{cc}
  \includegraphics[width=3.0in]{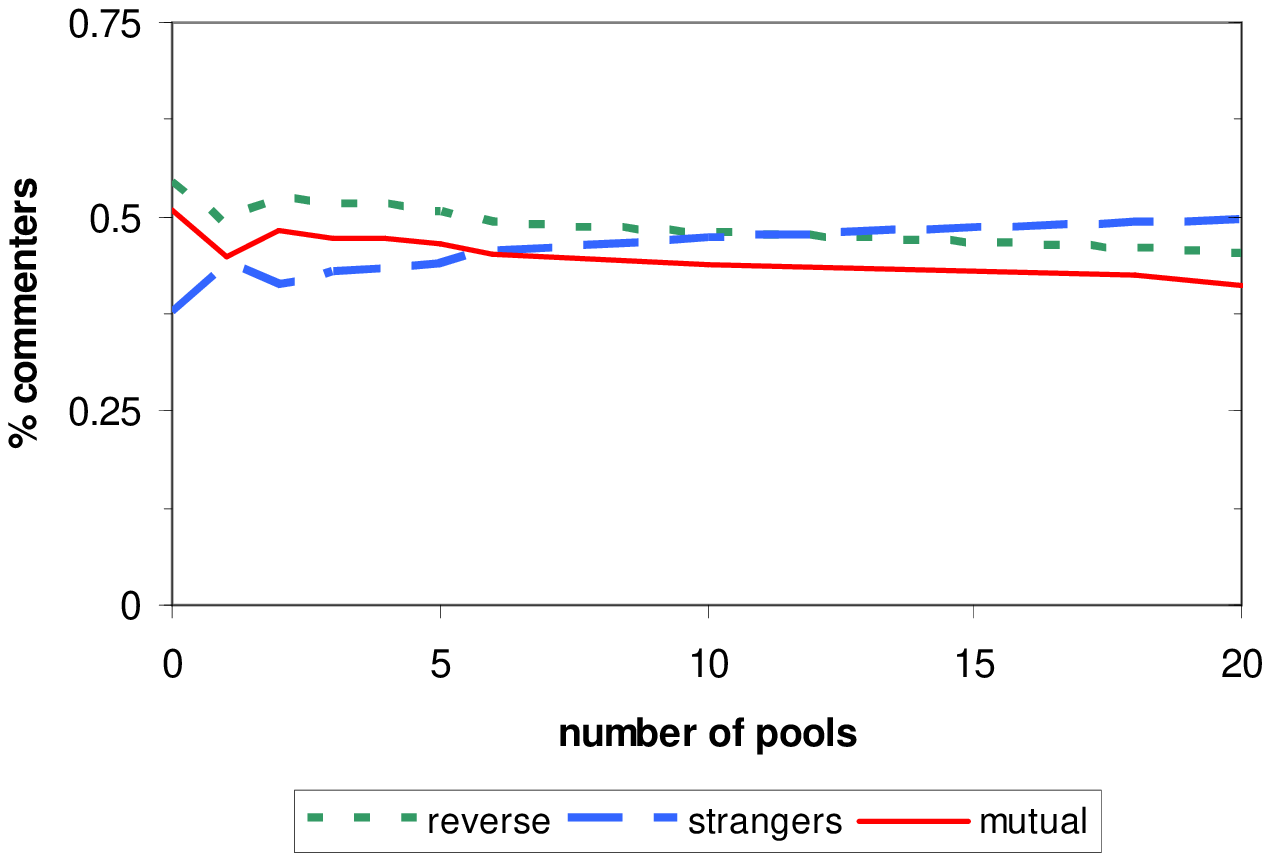} &

  \includegraphics[width=3.0in]{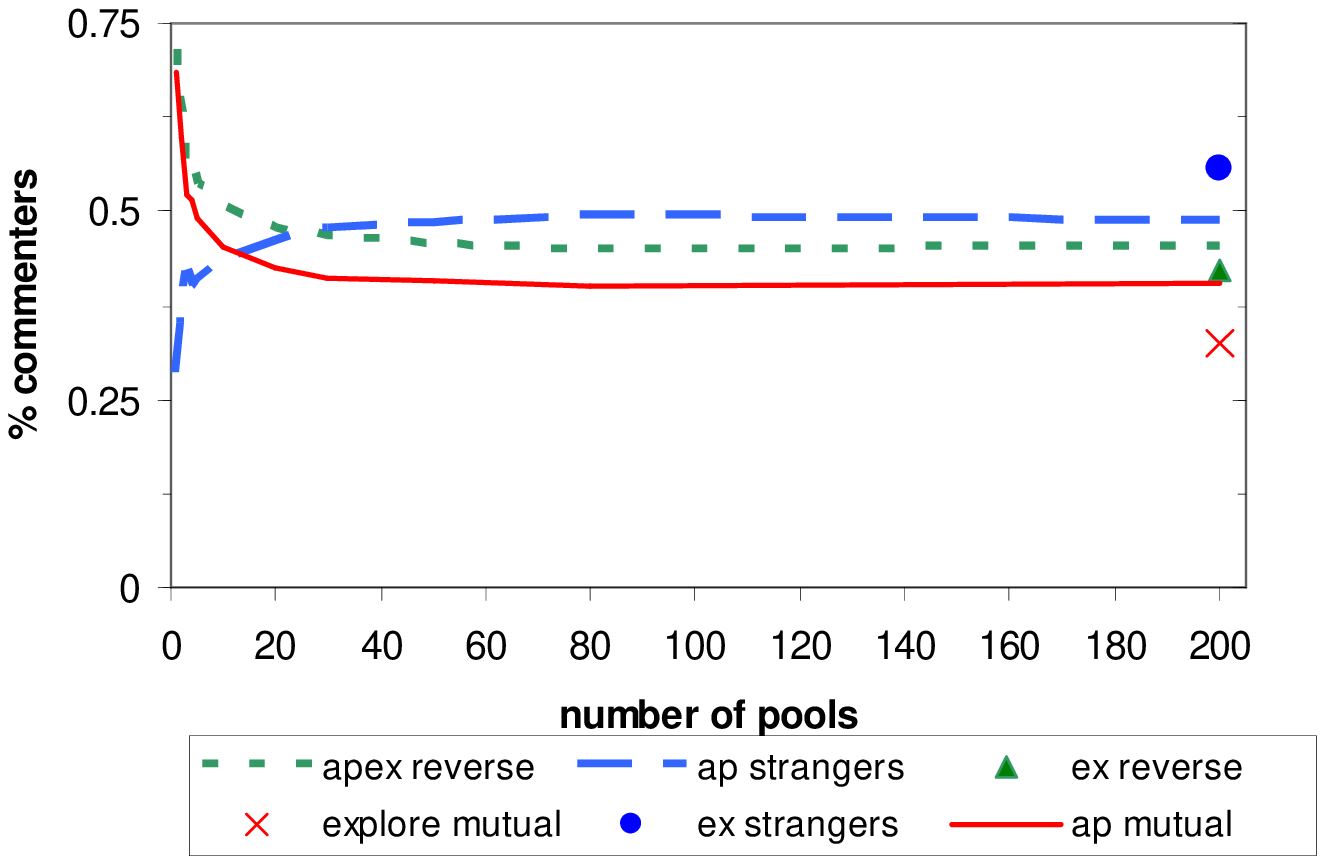} \\
 (a) \source{Random} & (b) \source{Apex} and \source{Explore}
\end{tabular}
  \caption{Proportion of comments
that came from the submitting user's reverse contacts, mutual
contacts and strangers vs the number of pools to which the image was
submitted for the three datasets}
  \label{fig:comments}
\end{figure}

\subsubsection{Social networks and comments}
\label{sec:socnetcomments}
We collected the names of users who commented on images in the
three datasets and compared them to the names of users in the image owners' social
networks. \figref{fig:comments} shows the proportion of comments
coming from owner's reverse contacts, mutual contacts and strangers
(users outside of the owner's social network). For \source{Random} images
(\figref{fig:comments}(a)) that were not
added to any pools, $55\%$ came from users who list the owner
as a contact, $51\%$ came from users who are mutual contacts of the
owner, while only $38\%$ came from users outside of the
owner's social network. As the image is posted to more and
more pools, its visibility to users outside of the owner's
social network grows. For \source{Random} images that have been
posted to 20 or more pools, only $41\%$ of the comments came from
mutual contacts, while the proportion of comments coming from
strangers grew to $49\%$.

These observations are even more pronounced for the \source{Apex}
images, shown in \figref{fig:comments}(b). For \source{Apex} images
that appear in only one pool (\source{Apex} itself), the share of
comments made by owner's mutual and reverse friends is
$69\%$ and $71\%$ respectively. Only $29\%$ of the comments came
from strangers. As the image gets shared with more groups, its
visibility to outsiders increases, up to a point. After an image has
been submitted to 30 groups, the share of the comments made by
mutual contacts drops to $41\%$, reverse contacts drops to $47\%$,
while the share of the comments coming from strangers grows to
$48\%$. The image's visibility to strangers does not appear to
increase by posting to additional groups. Sharing the image with 50
or more groups (up to 200) does not significantly change the
distribution of comments coming from contacts and strangers. This
seems to indicate that few of the groups are actively viewed (and
commented on) by users.\footnote{Groups such as the various 1-2-3
groups, Score Me or Delete Me groups require that the user view,
favorite or comment on other images in the pool before submitting
their own images. These groups are likely the ones driving most of
the traffic associated with posting images to groups.}

The symbols in \figref{fig:comments}(b) are for the \source{Explore}
images. We collected comments at the end of the tracking period,
after they have been publicly shared through the \emph{Explore} page. For
this set, $56\%$ of the comments come from strangers, far more than
for the other two sets, reflecting the \source{Explore} images'
greater public exposure. Still, about a third of the comments come
from mutual and $42\%$ from reverse contacts, showing that the
owner's social network is still active in commenting on and
presumably viewing the images.

In summary, we studied three groups of images: (a) images chosen
randomly from those uploaded on a specific day (\source{Random}
set), (b) images deemed by other photographers to be of exceptional
quality (\source{Apex} set) and (c) images chosen by Flickr's
Interestingness algorithm to be the best of those uploaded on a
specific day (\source{Explore} set). We analyzed a number of metrics
associated with these images --- the number of views, comments and
favorites they generated --- and studied the relationship of these
metrics to features such as the number of pools they were submitted
to, the number of tags associated with the images, and the size of
the users' social networks. \source{Explore} and \source{Apex}
images appear very similar on a number of metrics, despite the fact
that \source{Apex} images are months old (and presumably had more
time to be submitted to more pools or accumulate more tags) and very
different from the \source{Random} images. Judging by the size of
social networks, photographers from these two sets are also very
similar --- and distinct from the \source{Random} photographers.
This suggests that Interestingness algorithm does as good a job of
selecting good photographers as users do.\footnote{Surprisingly,
there is only a $10\%$ agreement between Interestingness and
photographers, because only 10\% of \source{Apex} images were
featured on the Explore page in the past.}

\section{Related Research}
\label{sec:related}
Many Web sites that provide information (or sell products or
services) use collaborative filtering technology to suggest relevant
documents (or products and services) to its users. Amazon and
Netflix, for example, use collaborative filtering to recommend new
books or movies to its users. Collaborative filtering-based
recommendation systems~\cite{Konstan97grouplens} try to find users
with similar interests by asking them to rate products and then
compare ratings to find users with similar opinions. Researchers in
the past have recognized that social networks present in the user
base of the recommender system can be induced from the explicit and
implicit declarations of user interest, and that these social
networks can in turn be used to make new
recommendations~\cite{ReferralWeb,perugini04}. Social media sites,
such as Digg and Flickr, are to the best of our knowledge the first systems to
allow users to explicitly construct social networks and use them for
information filtering. In addition to filtering, these social
networks can be used for information personalization, e.g.,
personalizing search results~\cite{Lerman07flickrsearch}.

Social navigation, a concept closely linked to collaborative filtering,
helps users evaluate the quality of information
by exposing information about the choices made by other
users ``through information traces left by
previous users for current users'' \cite{dieberger00}.
Exposing information about the choices made by others
has been has been shown~\cite{Salganik06,Wu07} to affect collective decision
making and lead to a large variance in popularity
of similar quality items. Unlike the present work, these research projects took into account only global information
about the preferences of others (similarly to the best seller lists and Top Ten albums). We believe
that by exposing local information about the choices of others within
your community, social browsing can lead to more effective information
filtering and collective decision making.

The proliferation of online networks~\cite{Garton97} has provided
interesting datasets about the behavior of large
groups and fueled interest in social networks from a variety of scientific disciplines.
Researchers have found that online social networks tend to augment existing
off-line relationships, and are often used to obtain emotional
support or expertise the user may lack in her offline
world~\cite{PewInternet}.
The rise of social networking sites such as LinkedIn, Friendster, MySpace,
Facebook, and many others, has introduced another interesting
domain to the study of computer-mediated interactions. These sites
are mainly used to link people who know each other offline: LinkedIn
is used to express professional relationships between colleagues,
while sites like Facebook and MySpace express friendship between college classmates and friends.
These sites are used to enhance offline interactions, by finding
recommendations for new jobs, dates~\cite{boyd04}, or simply keeping in touch with a
diverse group of existing offline friends~\cite{lampe06}, rather than
finding new online friends.
Social media sites are very different in nature from the purely social
networking sites mentioned above. Although some of the social
network connections on these sites express offline relationships, they play
but a minor role in the interactions found on these social media
sites. Instead, the public sharing of content and
metadata (e.g., tags) on sites like Digg and Flickr enables new social processing
applications, such as document evaluation and
ranking~\cite{Lerman07ic,Lerman07kdd}, information
discovery~\cite{Plangrasopchok07} and
personalization~\cite{Lerman07flickrsearch}.

\section{Conclusion}
\label{sec:conclusion}

Social media sites such as Flickr and Digg are on the leading edge of the
social Web revolution. These sites allow users to share and manage
content, participate in discussions, rate other people's activities, etc.
Importantly, they also
allow users to designate other users as friends or contacts. The
resulting social networks offer users new ways to interact with
information, through what we call social browsing or social
filtering.\footnote{These are part of a broader activity we call \emph{social information processing}.}

We studied social browsing on two popular social media sites: the photosharing site Flickr
and news aggregator Digg. We showed that social networks
form a basis for an effective information filtering system, suggesting to users the stories his friends
have found interesting. Users take advantage of these
recommendations simply by using the Friends interface to browse the
site. One of the implications of this finding,
elaborated in~\cite{Lerman07ic}, is that social browsing is the process by which
consensus, and the front (or Explore) page, emerge from the distributed opinions of many voters.
Rather than choosing complicated formulas to compensate for the
effect of social networks, Digg and Flickr could simply select
content based on whether it drew greater than expected
attention through views, favorites and comments.

We also claimed that social browsing is an important user interface
modality on Flickr. We offered two sources of evidence for this claim.
First, we showed that for the images produced by good photographers,
the views and favorites they receive correlate most strongly with
the number of reverse contacts the photographer has. We showed this
relationship directly by linking comments to the users in the
photographer's social network. Almost $3/4$ of the comments on the
images of good photographers, and $1/2$ of the \source{Random} ones,
come from other users within the photographer's social network.
Tags and pools are a less important ways to share images, except for \source{Random}
users, who do not have large social networks.
For both Digg and Flickr, while the front page (\emph{Explore} page on
Flickr) helps to
generate a large number of views and votes for the content (stories and images respectively), the
size of the submitter's social network appears to be the key to
promoting content to the front page.

Social media sites show that it is possible to exploit the activities of others
to solve hard information processing problems. We expect progress in this field to
continue to bring novel solutions to problems in information processing, personalization,
search and discovery.

\paragraph{Acknowledgements} The author wishes to acknowledge Laurie
Jones and Dipsy Kapoor for their invaluable help in gathering and
processing data from Flickr and Digg.
This research is based on work supported in part by the National Science
Foundation under Award Nos. IIS-0535182 and BCS-0527725.

\newpage

\bibliographystyle{plain}
\bibliography{../social,../lerman}

\end{document}